\documentclass[aps,twocolumn,pra,twoside,amssymb,amsmath]{revtex4}
\usepackage{graphicx}
\usepackage{amssymb}
\usepackage{amsfonts}
\usepackage{bbm}
\usepackage{amsmath}
\usepackage{lmodern}
\usepackage{mathrsfs}
\usepackage[dvipdfm,colorlinks=true, citecolor=blue, urlcolor=blue, linkcolor=blue]{hyperref}

\newcommand{\tr}{{\rm Tr}}
\newcommand{\idol}{\ensuremath{\mathbbm 1}}

\begin{document}
\title{Detecting and estimating coherence based on coherence witnesses}
\author{Zhao Ma$^{1,2}$}
\author{Zhou Zhang$^2$}
\author{Yue Dai$^2$}
\author{Yuli Dong$^2$}
\author{Chengjie Zhang$^{1,2}$}
\email{chengjie.zhang@gmail.com}
\affiliation{$^1$School of Physical Science and Technology, Ningbo University, Ningbo, 315211, China\\
$^2$School of Physical Science and Technology, Soochow University, Suzhou, 215006, China}

\begin{abstract}
Quantum coherence has wide-ranging applications from quantum thermodynamics to quantum metrology, quantum channel discrimination and even quantum biology. Thus,
detecting and quantifying coherence are two fundamental problems in quantum resource theory. Here, we introduce feasible methods to detect and estimate the coherence by constructing coherence witnesses for any finite-dimensional states. Our coherence witnesses detect coherent states by testing whether the expectation value of the witness is negative or not. Two typical coherence witnesses are proposed and discussed based on our witness-constructing method, which are also used to estimate the robustness of coherence, $l_1$-norm and $l_2$-norm of coherence measures.  Furthermore, we compare one of our coherence witness with a previously introduced witness, by proving that our witness is strictly stronger than that previous witness. We also present an application of coherence in a quantum metrology task, in which we estimate an unknown parameter by measuring our coherence witness.
\end{abstract}
\date{\today}

\maketitle

\section{Introduction}
Quantum coherence, a significant property in quantum theory, refers to the possibility of creating superpositions within a set of orthogonal states \cite{Johan,Mintert,Rev4}. It can be used to test the existence of quantumness in a single system and quantum correlations in composite systems, and has many applications \cite{g27,g28,Rev10,Rev19,Rev21,Rev22,Rev23,Rev26,Rev27,Rev20,exp2,exp3,exp4,exp5,Rev40,Rev14,Rev41,Rev43,Rev44,Rev45,Rev46,Rev47,Rev48,Rev17}.
Therefore, detecting and quantifying coherence become fundamental problems in quantum resource theory.


Consider a $d$-dimensional Hilbert space $\mathcal{H}$, a state is defined as an incoherent state under a reference basis $\{|i\rangle\}_{i = 1}^{d}$ if and only if its density matrix is diagonal under that reference basis \cite{Rev4}.  Therefore, one can represent all incoherent states as 
\begin{equation}
\delta = \sum_{i=1}^{d}   p_i|i\rangle\langle i|,
\end{equation}
where $0\leq p_i\leq1$ and $\sum_{i=1}^{d}   p_i=1$. We denote that $\mathbb{I}$ is the set of all incoherent states, and define the dephasing operation $\Delta$ as follows,
\begin{equation}\label{dephasing}
\Delta(\rho) = \sum_{i=1}^{d} |i\rangle\langle i|\rho|i\rangle\langle i|,
\end{equation}
which will be used to construct our coherence witnesses.

One of experimentally implementable ways to detect coherence is by measuring the expectation value of coherence witness.
Similar to entanglement witness \cite{Rev18,Rev49,Rev50,Rev51,Rev52,Rev53,Rev54,Rev55}, coherence witness was first introduced in Ref. \cite{Rev19}. It is defined that a Hermitian operator $W$ is a coherence witness if and only if $\tr(\delta W) \geq 0$ always holds for all incoherent states $\delta$. Therefore, if there exists a state $\rho$ such that $\tr(\rho W) < 0$, then $\rho$ must be a coherent state.
There are only a few papers discussed coherence witness \cite{Rev19,Rev21,Rev26,Rev27,Rev20}. Napoli and colleagues have proposed a lower bound of robustness of coherence by using coherence witness \cite{Rev19}.


To quantify the quantum coherence, many coherence measures have been proposed, such as distance-based coherence measures \cite{Rev4}, geometric measure of coherence \cite{Rev10}, robustness of coherence \cite{Rev19} and so on \cite{Rev40,Rev14,Rev41,Rev43,Rev44,Rev45,Rev46,Rev47,Rev48}.
According to the previous studies, a coherence measure $C$ should satisfy \cite{Rev4}: (C1) Non-negativity: $C(\rho)
\geq 0$ for any state $\rho$, and the equality holds if and only if $\rho$ is incoherent. (C2) Monotonicity:  $C(\Lambda[\rho]) \leq C(\rho)$ for any incoherent operation $\Lambda$ \cite{Rev4}, i.e. one cannot increase $C$  by using incoherent operations.
(C3) Strong monotonicity: $C$ does not increase on average under selective incoherent operations, $\sum_{i} q_i  C(\sigma_i) \leq C(\rho)$,
where $q_i= \tr(K_i\rho K_i^\dag)$, $\sigma_i$ are post-measurement states and satisfy $\sigma_i= (K_i\rho
K_i^\dag)/q_i$, and $\{K_i\}$ are incoherent Kraus operators. (C4) Convexity: $C$ is a convex function, i.e.,  $\sum_{i} p_i C(\rho_i) \geq C(\sum_{i} p_i
\rho_i)$ \cite{Rev4}. Some of coherence measures have analytical results. But there also exist coherence measures, which do not have analytical expressions yet. Thus, it is an interesting problem to estimate coherence measures, especially for experimental states.

The purpose of this work is two-fold. On the one hand, we present a general way of constructing coherence witnesses to detect coherent states. On the other hand, for estimating coherence, we get lower bounds of several coherence measures based on these coherence witnesses.  To this aim, we first introduce a simple and feasible method to construct coherence witnesses for any finite-dimensional states.  Based on our witness-constructing method, we propose two typical coherence witnesses. Moreover, we also use our coherence witnesses to estimate the robustness of coherence, $l_1$-norm and $l_2$-norm of coherence measures. Furthermore, we compare one of our coherence witness with a previously introduced witness, by proving that our witness is strictly stronger than that previous witness. We also present a quantum metrological task of coherence for estimating an unknown parameter by measuring our coherence witness.

\section{Detecting and estimating coherence based on coherence witnesses}
Before embarking on our main results, we first review some coherence measures in subsection A.
After that, we prove three theorems in subsection B. In theorem 1, by using any Hermitian matrix, we give a general constructing method of coherence witnesses with the diagonal elements being eliminated. In theorem 2, we propose the coherence witness $W_1$ by using density matrices, and we use it to  estimate the robustness of coherence and the $l_2$-norm of coherence measures. In theorem 3, we propose the coherence witness $W_2$, where its off-diagonal elements are complex numbers with modulus one. Moreover, based on the coherence witness $W_2$, we obtain a lower bound of the $l_1$-norm of coherence measure. Examples are presented in subsection C, including one example with real experimental results.

\subsection{Review of coherence measures}
A widely used coherence measure is the distance-based measure \cite{Rev4}
\begin{equation}
  C_{D}(\rho) = \min_{\delta \in \mathbb{I}}D(\rho, \delta),
\end{equation}
where $D(\rho, \delta)$ is the distance of $\rho$ to $\delta$, $\mathbb{I}$ is the set of all incoherent states, and $C_{D}(\rho)$ is the minimal distance. $C_{D}(\rho) = 0$ if and only if $\rho$ is incoherent.

The measure based on matrix norm is a kind of significant distance-based measure and the general form of its distance is $D(\rho,\delta) = \|\rho - \delta\|$ with $\|\cdot\|$ being some kind of matrix norm. One of the measures based on matrix norms  is the $l_p$ norm. The $l_p$ norm has the ideal property since it is directly related to the off-diagonal elements of the considered quantum state \cite{Rev25,Rev4}.

Among all the $l_p$ norms, $l_1$ norm is the most widely discussed. The $l_1$ norm of coherence can be denoted as:
\begin{equation}\label{W5}
  C_{l_{1}}(\rho) = \min_{\delta \in \mathbb{I}}\|\rho - \delta\|_{l_{1}} = \sum_{i,j}^{i \ne j}\ |\rho_{i,j}|.
\end{equation}
It is worth noticing that $l_1$ norm can be seen as the sum of the modulus of all off-diagonal elements in the matrix, so the function intuitively
 shows that whether there exists a non-zero off-diagonal element in the density matrix. Therefore, we can easily judge
  whether the state is coherent or not. 

We can also use other $ l_p$ norms to quantify the coherence. For instance, $ l_2$ norm can be applied to determine the coherence. The $ l_2$ norm of
coherence can be written as:
\begin{equation}
  C_{l_{2}}(\rho) = \min_{\delta \in \mathbb{I}}\|\rho - \delta\|_{l_{2}}^2 = \sum_{i \ne j}\ |\rho_{i,j}|^2.
\end{equation}

Besides, the robustness of coherence is defined in Ref. \cite{Rev19},
\begin{eqnarray}
C_{\mathcal{R}}(\rho)=\min_{\tau}\Big\{s\geq0 \Big| \frac{\rho+s\tau}{1+s}=:\delta\in \mathbb{I}\Big\},
\end{eqnarray}
where the minimum is taken over all possible quantum states $\tau$ such that its convexly mixed state with $\rho$ is an incoherent state.

\subsection{Detecting and estimating coherence based on coherence witness}

Based on the concept of coherence witness, we provide a general method of constructing coherence witnesses in the following.

\textit{Theorem 1.} For any Hermitian operator $A$, we can construct a coherence witness
 \begin{equation}\label{th1}
W = \Delta(A) - A,
\end{equation}
where $\Delta$ is the dephasing operation in Eq. (\ref{dephasing}).

\textit{Proof.---}
Because $A$ is a Hermitian operator, $W$  must be Hermitian.
For an arbitrary incoherent state $\delta = \sum_{i} p_{i}|i\rangle\langle i|$, we have
\begin{eqnarray}
\tr(\delta W) &=& \sum_{i}  p_i\langle i|\Delta(A)|i\rangle - \sum_{i}  p_i\langle i|A|i\rangle\nonumber\\
&=& \sum_{i}  p_i\langle i|\sum_{i'}|i'\rangle\langle i'|A|i'\rangle\langle i'|i\rangle - \sum_{i}  p_i\langle i|A|i\rangle.\nonumber
\end{eqnarray}
Since $\langle i'|i\rangle = \delta_{ii'}$, we can obtain
\begin{eqnarray}
&&\sum_{i} \ p_i\langle i|\sum_{i'}|i'\rangle\langle i'|A|i'\rangle\langle i'|i\rangle\nonumber\\
&=&\sum_{i}  p_i\langle i|\sum_{i'}|i'\rangle\langle i'|A|i'\rangle\delta_{ii'} \nonumber\\
&=&\sum_{i}   p_i\langle i|A|i\rangle.\nonumber
\end{eqnarray}
Thus,
\begin{equation}
    \tr(\delta W) = \sum_{i}   p_i\langle i|A|i\rangle - \sum_{i}  p_i\langle i|A|i\rangle= 0,
\end{equation}
which means $W$ is a coherence witness.  \hfill  $\blacksquare$

\textit{Remark.} There is another way to prove that $W$ in Eq. (\ref{th1}) is a coherence witness. In Ref. \cite{Rev19}, Napoli and colleagues have proved that a Hermitian operator $W$ is a coherence witness if and only if $\Delta(W)\geq0$, and it is easy to see the operator $W$ in Eq. (\ref{th1}) satisfies this condition, i.e., $\Delta(W)=\Delta[\Delta(A)]-\Delta(A)=0$ where we have used $\Delta[\Delta(A)]=\Delta(A)$.

Therefore, we have proved that the operator derived from our function is a coherence witness. This theorem suggests that we can construct a coherence
witness by finding an arbitrary Hermitian operator according to the density matrix we study, which will simplify the process and detect the coherence
more efficiently.

\textit{Theorem 2.} One can construct a coherence witness $W_1$ by choosing any density matrix $\sigma$ as the Hermitian operator $A$ in Eq. (\ref{th1}),
\begin{equation}\label{W1}
 W_1  =  \Delta(\sigma) -  \sigma,
\end{equation}
and this witness provides a lower bound of robustness of coherence for any quantum state $\rho$,
\begin{equation}\label{CR1}
    C_{\mathcal{R}} (\rho)\geq -\tr(\rho W_1).
\end{equation}
Specially, if $\sigma$ is a pure state $|\phi\rangle$,
\begin{equation}
    C_{\mathcal{R}} (\rho)\geq -\tr(\rho W_1)=F(\rho,|\phi\rangle)-\langle\Delta(|\phi\rangle\langle\phi|)\rangle,
\end{equation}
where the fidelity $F$ is defined as $F(\rho,|\phi\rangle)=\langle\phi|\rho|\phi\rangle$.
Furthermore, one can obtain
\begin{equation}
    C_{\mathcal{R}} (\rho)\geq C_{l_{2}} (\rho),
\end{equation}
for an arbitrary finite-dimensional quantum state $\rho$.

\textit{Proof.---}
Eq. (\ref{W1}) is obviously a coherence witness by choosing any density matrix $\sigma$ as the Hermitian operator $A$ in Eq. (\ref{th1}).

To prove inequality (\ref{CR1}), let us recall the inequality proved in Ref. \cite{Rev19}, i.e., if a coherence witness $ W\leq \idol$ then $C_{\mathcal{R}}(\rho)\geq -\tr(\rho W)$. From Eq. (\ref{W1}), one can see that
\begin{eqnarray}
\idol-W_1=\idol-\Delta(\sigma)+\sigma\geq0,
\end{eqnarray}
where the last inequality holds since $\sigma$ is a density matrix satisfying $\sigma\geq0$ and $\idol-\Delta(\sigma)\geq0$. Thus, $W_1\leq\idol$ holds, and we obtain inequality (\ref{CR1}).

Specially,
 if $\sigma$ is a pure state $|\phi\rangle$, then
\begin{eqnarray}
 C_{\mathcal{R}} (\rho)&\geq& -\tr(\rho W_1) \nonumber\\
                       &=& -\tr[\rho\Delta(|\phi\rangle\langle\phi|)]+\tr(\rho|\phi\rangle\langle\phi|)  \nonumber\\
                       &=& F(\rho,|\phi\rangle)-\langle\Delta(|\phi\rangle\langle\phi|)\rangle,
\end{eqnarray}
where the fidelity $F$ is defined as $F(\rho,|\phi\rangle)=\langle\phi|\rho|\phi\rangle$ and $\langle\Delta(|\phi\rangle\langle\phi|)\rangle:=\tr[\rho\Delta(|\phi\rangle\langle\phi|)]$.

Furthermore, if $\sigma$ is just $\rho$ itself, one can directly get that
\begin{eqnarray}
-\tr(\rho W_1) &=& -\tr\{[\Delta(\rho)-\rho]\rho\}\nonumber\\
         &=&\tr({\rho}^2) - \tr[\Delta(\rho)\rho]\nonumber\\
         &=&\sum_{i,j}  |\langle i|\rho|j\rangle|^2 - \sum_{i} |\langle i|\rho|i\rangle|^2\nonumber\\
         &=&\sum_{i \neq j} |\langle i|\rho|j\rangle|^2\nonumber\\
         &=&C_{l_{2}}(\rho),
\end{eqnarray}
where we have used the equations $\tr(\rho^2)=\sum_{i,j}  |\langle i|\rho|j\rangle|^2$ and $\tr[\Delta(\rho)\rho]=\sum_{i} |\langle i|\rho|i\rangle|^2$.
Combined with inequality (\ref{CR1}), one can see
\begin{equation}
    C_{\mathcal{R}} (\rho)\geq-\tr(\rho W_1)= C_{l_{2}} (\rho),
\end{equation}
which holds for any finite-dimensional quantum states.  \hfill  $\blacksquare$

\textit{Remark.} We substitute $A$ with $\sigma$ in theorem 1 to derive a specific coherence witness. Furthermore, we can directly construct a coherence witness by choosing a pure state $|\phi\rangle$ or $\rho$ itself as $\sigma$. If $\sigma=|\phi\rangle\langle\phi|$, our witness is experimentally accessible and related to fidelity. To realize a pure state $|\phi\rangle$ in experiments, we usually measure the fidelity between the target pure state $|\phi\rangle$ and the real experimental state $\rho$, and this fidelity can be directly used to obtain a lower bound of $C_{\mathcal{R}}$. We will provide an example using real experimental states.

In the following, we will relate the coherence witness to the $l_1$ norm and propose Theorem 3.

\textit{Theorem 3.} For any coherence witness $W  =  {\Delta(A) - A}$, where $A$ is an Hermitian matrix and the modulus of its off-diagonal elements is 1,
we can denote the coherence witness $W_2$ as

\begin{eqnarray}
W_2 = \left( \begin{array}{ccccc}
0 & e^{i\theta_{12}} & e^{i\theta_{13}} & \ldots  & e^{i\theta_{1d}} \\
e^{-i\theta_{12}} & 0 & e^{i\theta_{23}} & \ldots & e^{i\theta_{2d}} \\
e^{-i\theta_{13}} & e^{-i\theta_{23}} & 0 & \ldots & e^{i\theta_{3d}}\\
\vdots & \vdots & \vdots & \ddots   & \vdots \\
e^{-i\theta_{1d}} & e^{-i\theta_{2d}} & e^{-i\theta_{3d}}  & \ldots & 0
\end{array} \right),\label{W2}
\end{eqnarray}
where $\theta_{ij}$ can be arbitrary real values,
and this witness provides a lower bound of the $l_1$-norm of coherence
\begin{equation}\label{Cw2}
    C_{l_{1}}(\rho)\geq -\tr(\rho W_2).
\end{equation}

\begin{table*}
\begin{tabular}{c|ccccc}
\hline
\hline
  $\rho(N)$ & $6^a$ & $8^{a}$ & $8^{b}$ & $8^{c}$& $10^a$  \\
  \hline
  $F_{N}$  & $0.710(16)$ & $0.644(22)$ & $0.708(16)$ & $0.59(2)$ & $0.573(23)$ \\
  \hline
    $\langle P_{N}\rangle$  & $0.809(15)$ & $0.750(33)$ & $0.806(10)$ & $0.75(3)$ & $0.708(38)$ \\
  \hline
  $-\tr(\rho W_{3}) $ &0.210(16) &0.144(22) &0.208(16) &0.09(2) &0.073(23)  \\
  \hline
 $-\tr(\rho W_{1}) $ & $0.305(24)$ & $0.269(39)$ & $0.305(21)$ & $0.21(4)$ & $0.219(42)$ \\
  \hline
  \hline
\end{tabular}
\caption{The superscripts $a$, $b$, and $c$ denote experimental data from different sources of Refs. \cite{Nph1,Nph2,Nph3}. The fidelity $F_{N}$ is the result of overlap $\langle \phi |\rho|\phi\rangle$ with GHZ state $|\phi\rangle=1/\sqrt{2}(|H\rangle^{\otimes N}+|V\rangle^{\otimes N})$, and $\langle P_{N}\rangle$ is the expectation value of the project operator $P_N$. The values $-\tr(\rho W_{3})$ and $-\tr(\rho W_{1})$ are the lower bounds of robustness of coherence $C_{\mathcal{R}}$ based on Eqs. (\ref{Cw3}) and (\ref{CR1}), respectively. One can see that the lower bound $-\tr(\rho W_{1}) $ is always greater than the lower bound $-\tr(\rho W_{3}) $.  }\label{lab1}
\end{table*}

\textit{Proof.---}
Eq. (\ref{W2}) is obviously a coherence witness by choosing any Hermitian matrix, with the modulus of its off-diagonal elements being 1, as the Hermitian operator $A$ in Eq. (\ref{th1}).

Consider a $d$-dimensional density matrix
\begin{eqnarray}
\rho =  \left( \begin{array}{ccccc}
\rho_{11} & \rho_{12} & \rho_{13} & \ldots & \rho_{1d} \\
\rho_{21} & \rho_{22} & \rho_{23} & \ldots & \rho_{2d} \\
\rho_{31} & \rho_{32} & \rho_{33} & \ldots & \rho_{3d} \\
\vdots & \vdots & \vdots & \ddots & \vdots \\
\rho_{d1} & \rho_{d2} & \rho_{d3} & \ldots & \rho_{dd}
\end{array} \right),
\end{eqnarray}
we have
\begin{eqnarray}
\tr(\rho W_2)  &=& \sum_{j < k} \rho_{jk}e^{-i\theta_{jk}}  + \sum_{j > k} \rho_{jk}e^{i\theta_{kj}} \nonumber\\
&=& \sum_{j < k} (\rho_{jk}e^{-i\theta_{jk}} + \rho_{jk}^{*}e^{i\theta_{jk}}).
\end{eqnarray}
Since the density matrix is a Hermitian matrix, thus $\rho_{jk}  =  \rho_{kj}^\ast$, and we can get that $\rho_{jk} e^{-i\theta_{jk}} = (\rho_{kj}
e^{i\theta_{jk}})^\ast$. Let $\rho_{jk} = |\rho_{jk}|e^{i\psi_{jk}}$, and we will have
\begin{eqnarray}
-\tr(\rho W_2) &=& -\sum_{j < k} |\rho_{jk}|(e^{i(\psi_{jk}-\theta_{jk})} + e^{-i(\psi_{jk}-\theta_{jk})})\nonumber\\
 &\leq& 2 \sum_{j < k} |\rho_{jk}| = C_{l_1}.
\end{eqnarray}
Thus, inequality (\ref{Cw2}) has been proved.  \hfill  $\blacksquare$

\textit{Remark.} We can select a specific value $\theta_{jk}=\psi_{jk}$ in order to make the phase $\psi_{jk}$ vanish. Thus only the modulus of density matrix' off-diagonal elements will preserve, and the equality in (\ref{Cw2}) holds. Generally, the result of witness $W_2$ provides a lower bound of $C_{l_1}$. In subsection C, we will provide an example to show that $-\tr(\rho W_2)$ is the lower bound of $C_{l_1}$.

In Ref. \cite{Rev26}, a coherence witness was presented as a coherence bound similar to the entanglement witness:
\begin{equation}\label{W3}
    W_3=|\lambda_{\mathrm{max}}|^2\idol-|\phi\rangle\langle\phi|,
\end{equation}
where $|\phi\rangle=\sum_{i=1}^d \lambda_{i}|i\rangle$ is an arbitrary
$d$-dimensional pure state with $|\lambda_{\mathrm{max}}|=\max_i\{|\lambda_{i}|\}$. For all the incoherent states $\delta=\sum_i p_i |i\rangle\langle i|$,
we have $\tr(W_3\delta)=|\lambda_{\mathrm{max}}|^2-\sum_i p_i |\lambda_i|^2\geq0$. If there exists a state
$\rho$ such that $\tr(W_{3}\rho)<0$, the state $\rho$ must be a coherent state. Similar to $W_1$, one can easily prove that $W_3\leq\idol$ and then we have
\begin{equation}\label{Cw3}
  C_{\mathcal{R}} (\rho)\geq -\tr(\rho W_3).
\end{equation}
Then we can compare $W_1$ with $W_3$ in the following.

\textit{Proposition 1.} The coherence witness $W_1$ in Eq. (\ref{W1}) with $\sigma = |\phi\rangle\langle\phi|$  is strictly stronger than $W_3$ in Eq.
(\ref{W3}).

\textit{Proof.---}
Consider a $d$-dimensional witness $W_1$ and $W_3$.
Since $\sigma$ has be chosen as $\sigma = |\phi\rangle\langle\phi|$,
\begin{eqnarray}
\Delta(\sigma) =\mathrm{diag}\{|\lambda_1|^2,|\lambda_2|^2,\cdots,|\lambda_d|^2\}.
\end{eqnarray}

In order to show that $W_1$ is strictly stronger than $W_3$, we should prove $W_1\leq W_3$. We can obtain
\begin{eqnarray}
 &&W_3 - W_1\nonumber\\
 &=& |\lambda_{\mathrm{max}}|^2\idol -\Delta(\sigma)\nonumber\\
 &=& \mathrm{diag}\{|\lambda_{\mathrm{max}}|^2-|\lambda_1|^2,\cdots,|\lambda_{\mathrm{max}}|^2-|\lambda_d|^2\}\nonumber\\
 &\geq&0.
\end{eqnarray}
This means for arbitrary state $\rho$ we have
\begin{eqnarray}
\tr[\rho(W_3 - W_1)]
= \sum_{i=1}^d |\lambda_i|^2(|\lambda_{\mathrm{max}}|^2 - |\lambda_i|^2)
\geq 0.
\end{eqnarray}
Therefore, $\tr(W_3\rho)\geq\tr(W_1\rho)$ holds. On the one hand, there may exist a state $\rho'$ for which $\tr(W_3\rho')\geq0$ but $\tr(W_1\rho')<0$, i.e.,  the coherence of $\rho'$ can be detected by $W_1$ but cannot be detected by $W_3$. On the other hand, if a state $\rho''$ is detected by $W_3$, i.e., $\tr(W_3\rho'')<0$, one can obtain that $0>\tr(W_3\rho'')\geq\tr(W_1\rho'')$ which means $\rho''$ is also detected by $W_1$. \hfill  $\blacksquare$

\textit{Remark.} According to the result above, we can conclude that the coherence of the state which can be detected by $W_3$ can also be detected by $W_1$ while the opposite may not be true. Therefore, our witness is strictly stronger than $W_3$ for pure states.

\subsection{Examples}
\textit{Example 1.---} We adopt three experimental results of multi-photon entanglement \cite{Nph1,Nph2,Nph3} to demonstrate the witness $W_{1}$ is strictly stronger than $W_{3}$ for the experimentally realized mixed states. These experiments create Greenberger-Horne-Zeilinger states (GHZ) with $6-10$ photons and  calculate the probability of  $P^{N}$ as well as the fidelity of their prepared GHZ state, where $P^{N}=(|H\rangle\langle H|)^{\otimes N}+(|V\rangle\langle V|)^{\otimes N}$ represents the project operator which denotes the
population of the $|H\rangle^{\otimes N}$ and $|V\rangle^{\otimes N}$ components of the GHZ state. Let $|H\rangle$ ($|V\rangle$) be $|0\rangle$ ($|1\rangle$), and we choose the computational basis as the reference basis. When $|\phi \rangle$ in Eq. $(\ref{W3})$ is selected as $N$-GHZ state, the lower bound of robustness of coherence in (\ref{Cw3})  can then be directly calculated as
\begin{equation}
  C_{\mathcal{R}} (\rho)\geq -\tr(\rho W_{3} )=  F_{N}-\frac12,
\end{equation}
where $F_{N}=\langle \mathrm{GHZ}^{N}|\rho|\mathrm{GHZ}^{N}\rangle$ is the fidelity of the experimentally produced state $\rho$ with the ideal $N$-GHZ state. When $\sigma$ in Eq. $(\ref{W1})$ is chosen as $N$-GHZ state, we can get another lower bound of robustness of coherence,
\begin{eqnarray}
C_{\mathcal{R}} (\rho)&\geq& -\tr(\rho W_{1} ) \nonumber\\
 &=& -\tr[\rho(\Delta(|\mathrm{GHZ}^{N}\rangle \langle \mathrm{GHZ}^{N}|)-|\mathrm{GHZ}^{N}\rangle \langle \mathrm{GHZ}^{N}|)] \nonumber\\
&=&-\tr\Big(\rho\frac{P^{N}}2-\rho|\mathrm{GHZ}^{N}\rangle \langle \mathrm{GHZ}^{N}|\Big) \nonumber \\
 &=&   F_{N}-\frac{\langle P^{N}\rangle}2.
\end{eqnarray}
From Table \ref{lab1}, we can see that the lower bound $-\tr(\rho W_{1}) $ is always greater than the lower bound $-\tr(\rho W_{3}) $  when $N$ is 6,8 and 10 photons, so it is experimentally demonstrated that the witness $W_{1}$ is strictly stronger than $W_{3}$.

\begin{figure}
\includegraphics[scale=0.9]{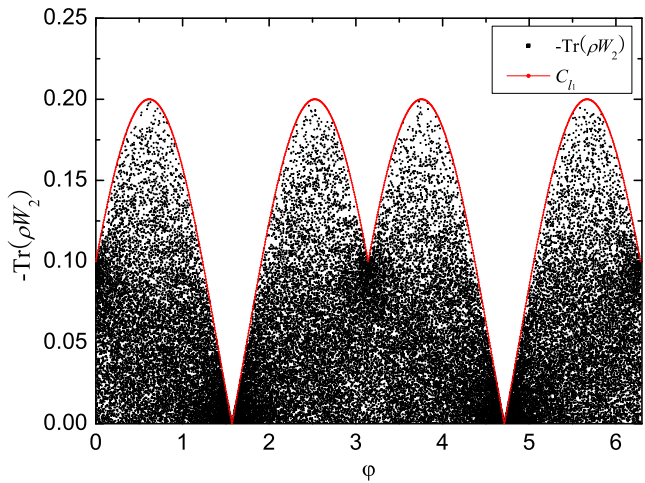}
\caption{The red solid line represents $C_{l_{1}}$,  and the black dots are $-\tr(\rho W_2)$ for the density matrix in Eq. (\ref{Eqp}), where $p = 0.1$, $\phi_1 = 0.2$, and $\phi_2 = 0.3$. The negative values of $-\tr(\rho W_2)$ are ignored for the ineffectivity of $W_2$ under such situations.}\label{1}
\end{figure}

\textit{Example 2.---}To demonstrate the lower bound of $C_{l_1}$ in Eq. (\ref{Cw2}), we consider a 3-dimensional state,
\begin{equation}\label{Eqp}
\rho = p|\Phi\rangle\langle\Phi| + (1-p)\frac{\idol}{3},
\end{equation}
where $p$ is the probability, and $|\Phi\rangle = (\cos\varphi/\sqrt{2})|1\rangle + (\cos\varphi e^{i\phi_1}/\sqrt{2})|2\rangle + \sin\varphi e^{i\phi_2}|3\rangle$.

From Theorem 3, we can generate many coherence witnesses $W_2$ by changing the phases of the off-diagonal elements in the matrix. Thus we randomly
select the phases of the elements, and the results of $-\tr(\rho W_2)$ are presented in Fig. \ref{1}, which intuitively reveal that $-\tr(\rho W_2)$ is
strictly smaller than $C_{l_1}$,
\begin{equation}
C_{l_1}(\rho) = p\big(\cos^2{\varphi} + 2\sqrt{2}|\cos{\varphi}\sin{\varphi}|\big).
\end{equation}

Specially, the phase of each element of $W_2$ is chosen equally to that of $\rho$, then the $-\tr(\rho W_2)$ reaches the maximum and equals to $C_{l_1}$. The conclusion indicates that for a incoherent state $\rho$, the $-\tr(\rho W_2)$ must be zero because there is no possibility for it to be above $C_{l_1}$. Also, if $C_{l_1}$ is not zero, $\tr(\rho W_2)$ will be negative, which suggests that the state is coherent.


\section{Application in quantum metrology}
There are many applications of quantum coherence. For instance, quantum coherence is of great usefulness in quantum thermodynamics, and some properties of coherence operations can be related to the basic laws of thermodynamics \cite{thermodynamics1,thermodynamics2}. Furthermore, quantum coherence can play a fundamental role in certain quantum metrology tasks, such as the phase discrimination game \cite{Rev19}.

Here we present a coherence application in a quantum metrology task.  In Ref. \cite{g27}, a simple example of quantum metrology has been presented, in which quantum coherence plays a fundamental role. However, the example only concerned single-qubit pure states, we now provide a general result for arbitrary finite dimensional quantum states.

Consider a $d$-dimensional Hibert space, let $H$ be a non-degenerate Hamiltonian,
 \begin{equation}\label{nondeH}
 H=\sum_{i=0}^{d-1}E_i|i\rangle\langle i|,
 \end{equation}
where $\{E_i\}$ and $\{|i\rangle\}$ are its eigenvalues and eigenvectors, respectively, and $E_i\neq E_j$ when $i \neq j$. We can choose its eigenvectors $|i\rangle$ as the reference basis of coherence. Let us focus on the following parameter estimation task.

\begin{figure}
\includegraphics[scale=0.9]{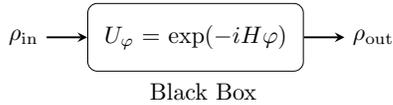}
\caption{The Black Box implements an unitary $U_{\varphi}=\exp(-iH \varphi)$ on an input quantum state $\rho_{\mathrm{in}}$ with its output state being $\rho_{\mathrm{out}}=U_{\varphi}\rho_{\mathrm{in}} U_{\varphi}^\dag$. Suppose that the full information of non-degenerate Hamiltonian $H$ is already known. The quantum metrology task is to estimate the unknown parameter $\varphi$ by measuring the output state $\rho_{\mathrm{out}}$. }\label{2}
\end{figure}

In Fig. \ref{2}, a black box implements an unitary $U_{\varphi}=\exp(-iH \varphi)$ on an input quantum state $\rho_{\mathrm{in}}$ with its output state being $\rho_{\mathrm{out}}=U_{\varphi}\rho_{\mathrm{in}} U_{\varphi}^\dag$. Suppose that we have obtained the full information of the Hamiltonian $H$ Eq. (\ref{nondeH})  in $U_\varphi$, but know nothing about the parameter $\varphi$. Can we gain the information of the parameter $\varphi$ by measuring the output state $\rho_{\mathrm{out}}$?  If $\rho_{\mathrm{out}}$ is independent of $\varphi$, it cannot be used for estimating $\varphi$. We can express $\rho_{\mathrm{in}}$ under the reference basis,
\begin{equation}
\rho_{\mathrm{in}}=\sum_{i,j}\rho_{i,j}|i\rangle\langle j|,
\end{equation}
with $\rho_{i,j}=\langle i|\rho_{\mathrm{in}}|j\rangle$. Thus,
\begin{eqnarray}
\rho_{\mathrm{out}}&=&U_{\varphi}\rho_{\mathrm{in}} U_{\varphi}^\dag    \nonumber \\
&=&\sum_{i\neq j}\rho_{i,j} e^{-i(E_i-E_j)\varphi}|i\rangle\langle j|+\sum_i \rho_{i,i}|i\rangle\langle i|. \label{rhoout}
\end{eqnarray}
From Eq. (\ref{rhoout}), one can see that $\rho_{\mathrm{out}}$ is dependent of $\varphi$ if and only if there exist $i_0$ and $j_0$ with $i_0\neq j_0$ such that $\rho_{i_0,j_0}\neq 0$,  since $\rho_{i_0,j_0} e^{-i(E_{i_0}-E_{j_0})\varphi}\neq 0$. Therefore, we can obtain the following proposition.

\textit{Proposition 2.---}  Let $H$ be a non-degenerate Hamiltonian, and we choose its eigenvectors $|i\rangle$ as the reference basis of coherence. An input state $\rho_{\mathrm{in}}$ can be used to estimate the unknown pamameter $\varphi$ in the Black Box in Fig. \ref{2}, if and only if $\rho_{\mathrm{in}}$ has nonzero coherence under the reference basis. Furthermore, one can also obtain that $\rho_{\mathrm{out}}$ has nonzero coherence under the reference basis if and only if $\rho_{\mathrm{in}}$ has nonzero coherence as well.

\textit{Example 3.---} Based on Proposition 2, we choose the maximally coherent state $|\phi^+\rangle=\frac{1}{\sqrt{d}}\sum_{i=0}^{d-1}|i\rangle$ as the input state $\rho_{\mathrm{in}}$, where the reference basis $\{|i\rangle\}$ are eigenvectors of non-degenerate Hamiltonian $H$ in Eq. (\ref{nondeH}). Thus,
\begin{eqnarray}
\rho_{\mathrm{out}}=U_{\varphi}|\phi^+\rangle\langle\phi^+|U_{\varphi}^\dag
=\frac{1}{d}\sum_{i,j}e^{-i(E_i-E_j)\varphi}|i\rangle\langle j|.
\end{eqnarray}
One can construct a coherence witness to detect the coherence of $\rho_{\mathrm{out}}$.

Suppose that $E_{\mathrm{max}}$ ($E_{\mathrm{min}}$) is the maximal (minimal) eigenvalues of $H$, with the corresponding eigenvector being $|E_{\mathrm{max}}\rangle$ ($|E_{\mathrm{min}}\rangle$). We select $A=|E_{\mathrm{max}}\rangle\langle E_{\mathrm{max}}|+|E_{\mathrm{max}}\rangle\langle E_{\mathrm{min}}|+|E_{\mathrm{min}}\rangle\langle E_{\mathrm{max}}|+|E_{\mathrm{min}}\rangle\langle E_{\mathrm{min}}|$. Therefore,
\begin{eqnarray}
W&=&-|E_{\mathrm{max}}\rangle\langle E_{\mathrm{min}}|-|E_{\mathrm{min}}\rangle\langle E_{\mathrm{max}}|,\\
\tr(W\rho_{\mathrm{out}})&=&-\frac{1}{d}[e^{-i(E_{\mathrm{max}}-E_{\mathrm{min}})\varphi}+e^{i(E_{\mathrm{max}}-E_{\mathrm{min}})\varphi}]  \nonumber\\
&=&-\frac{2}{d} \cos[(E_{\mathrm{max}}-E_{\mathrm{min}})\varphi].
\end{eqnarray}
One can preform the above measurement with repeating $n$ times, then the mean square error $(\Delta\varphi)^2$ is bounded by the quantum Cram\'er-Rao bound \cite{CR}.

\section{Discussions and conclusions}
In Ref.\cite{Rev27}, another coherence witness has been presented. Any $d$-dimensional state can be written as $\rho = \frac{\idol}{d} + \frac{1}{2}\sum_{j=1} ^{d-1}\sum_{k=j+1}^{d}(b_{s}^{jk}\sigma_{s}^{jk} + b_{a}^{jk}\sigma_{a}^{jk}) + \frac{1}{2}\sum_{l=2}^{d}b^{l}\sigma^{l}$ \cite{Rev33}. Similarly, the general form of a coherence witness is
\begin{equation}
W_4 = w_{0}\idol + \sum_{jk} (w_{s}^{jk}\sigma_{s}^{jk} + w_{a}^{jk}\sigma_{a}^{jk}) + \sum_{l=2}^{d}w^{l}\sigma^{l},
\end{equation}
where $b_{s,a}^{jk} = \tr(\rho \sigma_{s,a}^{jk})$, $b^l = \tr(\rho \sigma^l)$. $w_{s,a}^{jk}$, $w_0$, and $w^l$ are real numbers. $\sigma_{s}^{jk} = |j\rangle\langle k| + |k\rangle\langle j|$, $\sigma_{a}^{jk} = -i|j\rangle\langle k| + i|k\rangle\langle j|$ and $\sigma^{l} = \sqrt{\frac{2}{l(l+1)}}(\sum_{j=1}^{l}|j\rangle\langle j|-l|l\rangle\langle l|)$ are the Gell-Mann matrices \cite{Rev33}.

According to its definition, $\tr(\rho W_4) = 0$ for all $\rho \in \mathbb{I}$. Hence $w_0 = w^l = 0$. Then we normalize it and the general form of $W_4$ is simplified as
\begin{equation}
W_4 = \sum_{jk} (\cos\theta_{jk}\sigma_{s}^{jk} + \sin\theta_{jk}\sigma_{a}^{jk}),
\end{equation}
where $\theta \in [0, 2\pi)$.

Thus
\begin{equation}
\tr(\rho W_4) = \sum_{jk}(\cos\theta_{jk}b_{s}^{jk} + \sin\theta_{jk}b_{a}^{jk}).
\end{equation}
Let $b_{a}^{jk} = |b^{jk}|\cos\psi_{jk}$ and $b_{s}^{jk} = |b^{jk}|\sin\psi_{jk}$, so we can rewrite $\tr(\rho W_4)$ as $\tr(\rho W_4) = \sum_{jk}|b_{jk}|(\cos\theta_{jk}\cos\psi_{jk} + \sin\theta_{jk}\sin\psi_{jk}) = \sum_{jk}|b_{jk}|\cos(\theta_{jk} - \psi_{jk})$. The witness $W_4$ is related to $l_1$ norm, and the equation $C_{l_{1}}(\rho) = \sum_{jk} \sqrt{(b_{s}^{jk})^2 + (b_{a}^{jk})^2}$ embodies all the properties of the $l_1$ norm. $\tr(\rho W_4) \leq C_{l_1}$ always holds, and "=" will be achieved if and only if $\theta_{jk} = \psi_{jk}$. We can find this result similar to ours derived from $W_2$.

In conclusion, we have presented a general way of constructing a coherence witness. Besides, we have provided two examples of constructing specific
coherence witnesses related to some coherence measures. These coherence witnesses are very efficient and can be adopted to simplify the process
of detecting coherence of certain states. The coherence witness can be used in a wide range of fields, such as quantum thermodynamics, quantum
algorithms, quantum metrology, quantum channel discrimination and witnessing quantum correlations \cite{g27,g28}. However, our method of deriving a coherence
witness still has some deficiencies. Because we have just  related the coherence witnesses to the robustness of coherence, $l_1$ and $l_2$ norms of coherence, other coherence measures still remain to be studied. We believe that our work will increase the efficiency of detecting and estimating the coherence and contribute to the research of quantum coherence.

\section*{ACKNOWLEDGMENTS}
This work is supported by the National Natural Science Foundation of China (Grant No. 11734015), and K.C. Wong Magna Fund in Ningbo University.

\end{document}